\documentclass[aps,pra,twocolumn,a4paper,showpacs,superscriptaddress,amsmath,amssymb]{revtex4-1}

\usepackage{graphicx}
\usepackage{hyperref}
\usepackage{graphicx}
\usepackage{textcomp}
\usepackage[normalem]{ulem}
\usepackage{color}
\usepackage{pgfplots}
\pgfplotsset{compat=newest}
\usepackage{siunitx}
\usepgfplotslibrary{groupplots}
\usetikzlibrary{matrix,positioning}

\usepackage{shellesc}
\usepackage{tikz} 
\usetikzlibrary{circuits.ee.IEC}
\usetikzlibrary{shapes}
\usetikzlibrary{external}
\usepackage{rotating}
\definecolor{utkarminrot}{cmyk}{0.35,1,0.7,0.1} 
\definecolor{utgold}{cmyk}{0.3,0.3,0.6,0.1} 
\definecolor{utanthrazit}{cmyk}{0.3,0,0,0.85} 
\definecolor{utblaulila}{cmyk}{0.7,0.5,0,0.35} 
\definecolor{utblau}{cmyk}{1,0.5,0.1,0} 
\definecolor{uthellblau}{cmyk}{0.6,0,0,0.2} 
\definecolor{uttuerkis}{cmyk}{0.5,0,0.5,0.1} 
\definecolor{utgruen}{cmyk}{0.5,0,0.8,0.2} 
\definecolor{utdunkelgruen}{cmyk}{0.7,0,1,0.5} 
\definecolor{utrot}{cmyk}{0.2,0.8,0.8,0} 
\definecolor{utrosa}{cmyk}{0.2,0.6,0,0.2} 
\definecolor{uthellbraun}{cmyk}{0.2,0.3,0.3,0.2} 
\definecolor{utocker}{cmyk}{0.2,0.3,0.65,0} 
\definecolor{utorange}{cmyk}{0.1,0.4,1,0.1} 
\definecolor{utbraun}{cmyk}{0.3,0.5,0.7,0.3}

\usetikzlibrary{shapes.multipart,arrows,calc}

\usetikzlibrary{shapes,snakes}

\tikzset{fnode/.style={rectangle split, rectangle split part align={left,left,left,left,left}, rectangle split parts=5, draw, minimum width =2.75cm, rounded corners}}

\tikzset{fnode1/.style={rectangle, rectangle split part align={left}, draw, minimum width =1.25cm, rounded corners}}
\tikzset{fnode2/.style={rectangle split, rectangle split part align={left,left}, rectangle split parts=2, draw, minimum width =2.5cm, rounded corners}}

\tikzset{label style/.style={draw, rounded corners}}

\begin{document}

	\title{Absolute frequency measurement of rubidium 5S-6P transitions}

	\author{Conny Glaser}\email{conny.glaser@uni-tuebingen.de}
	\affiliation{Center for Quantum Science, Physikalisches Institut, Eberhard Karls Universit\"at
		T\"ubingen, Auf der Morgenstelle 14, D-72076 T\"ubingen,
		Germany}
		\author{Florian Karlewski}
	\affiliation{HighFinesse GmbH, W\"ohrdstra{\ss}e 4, D-72072 T\"ubingen, Germany}
		\author{Jens Grimmel}
	\affiliation{Center for Quantum Science, Physikalisches Institut, Eberhard Karls Universit\"at
	T\"ubingen, Auf der Morgenstelle 14, D-72076 T\"ubingen,
	Germany}	
			\author{Manuel Kaiser}
\affiliation{Center for Quantum Science, Physikalisches Institut, Eberhard Karls Universit\"at
	T\"ubingen, Auf der Morgenstelle 14, D-72076 T\"ubingen,
	Germany}	
			\author{Andreas G\"unther}
	\affiliation{Center for Quantum Science, Physikalisches Institut, Eberhard Karls Universit\"at
	T\"ubingen, Auf der Morgenstelle 14, D-72076 T\"ubingen,
	Germany}
		\author{Helge Hattermann}
	\affiliation{Center for Quantum Science, Physikalisches Institut, Eberhard Karls Universit\"at
		T\"ubingen, Auf der Morgenstelle 14, D-72076 T\"ubingen, Germany}
	\author{J{\'o}zsef Fort\'agh}
	\affiliation{Center for Quantum Science, Physikalisches Institut, Eberhard Karls Universit\"at
		T\"ubingen, Auf der Morgenstelle 14, D-72076 T\"ubingen,
		Germany}
	\begin{abstract}
			We report on measurements on the 5S-6P rubidium transition frequencies for rubidium isotopes with an absolute uncertainty of better than \SI{450}{kHz} for the 5S $\rightarrow$ 6P$_{1/2}$ transition and \SI{20}{kHz} for the 5S $\rightarrow$ 6P$_{3/2}$ transition, achieved by saturation absorption spectroscopy. From the results we derive the hyperfine splitting with an accuracy of \SI{460}{kHz} and \SI{30}{kHz}, respectively. We also verify the literature values for the isotope shifts as well as magnetic dipole constant and the electric quadrupole constant.
		%
		%
	\end{abstract}
	\pacs{}
	
	\maketitle
	
	\section*{Introduction}
The advent of optical frequency combs has revolutionized the world of high precision spectroscopy and has enabled measurements of atomic transition frequencies with exceptional accuracy \cite{udem2002optical, Udem2009, Cundiff2003}. 
Precise knowledge of these frequencies facilitates better calculations of atomic models and the derivation of more accurate values of physical quantities such as the Lamb shift \cite{huang2018frequency}, hyperfine structure constants \cite{Lee2015}, magnetic dipole or electric quadrupole constant.
Furthermore, precise know\-ledge of these values is desirable to experimentally investigate novel ways of manipulating atomic states,
%
such as the coherent excitation of Rydberg atoms for quantum information processing \cite{saffman2010quantum, levine2018high}, generation of atomic memories \cite{patton2013ultrafast} or implementing novel quantum gates \cite{sarkany2015long}.

For quantum information purposes rubidium Rydberg atoms are a widely used species. A common way to excite Rubidium atoms from the 5S ground state to Rydberg states with high principal quantum numbers n is via a three level ladder scheme 5S$\rightarrow$5P$\rightarrow$nS or nD using a pair of lasers with wavelengths 780nm and 480nm.
This scheme, however, commonly relies on the generation of \SI{480}{\nano \metre} light by frequency-doubling a \SI{960}{\nano \metre} laser, which limits the available laser power \cite{mack2011measurement}. \\
A promising alternative is the excitation using the 6P state as intermediate state \cite{simonelli2017deexcitation,gutierrez2017experimental}. 
Due to the five times larger lifetime $\tau$ = \SI{121}{\nano \second} compared to the 5P state \cite{Gomez04}, dephasing during the excitation is reduced and the coherence time of this transition is expected to be larger \cite{sarkany2015long, sarkany2018faithful}.
%
%
The commercial availability of \SI{420}{\nano \metre} ECDL lasers renders the Rydberg excitation (ladder) scheme via the 6P intermediate state a viable alternative to the commonly used excitation state via the 5P state. 
Additionally, the light driving the 6P\,$\rightarrow$\,nS transition at \SI{1016}{\nano \metre} can easily be generated with high power using external cavity diode lasers (ECDL) and allows high Rabi frequencies in the excitation to Ryd\-berg states.

\begin{figure}[tbp]
	\centerline{
		\includegraphics[width=\linewidth]{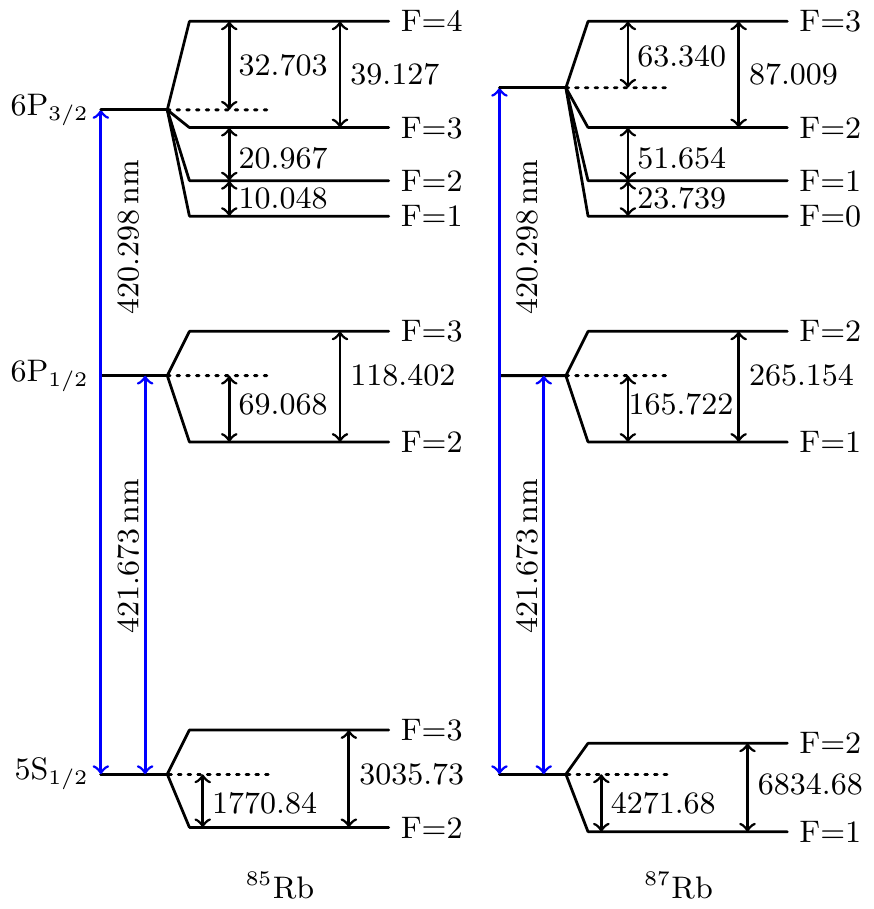}}
	\caption{(Color online) Level scheme and hyperfine splitting (in MHz) for the 5S$_{1/2}$ and 6P manifold of $^{85}$Rb (left) and $^{87}$Rb (right). Drawing not to scale.}
	\label{abstaende}
\end{figure}

While the transition frequencies for the excitation scheme 5S\,$\rightarrow$\,5P\,$\rightarrow$\,nS, nD are known to \SI{6}{\kilo\Hz} accuracy \cite{steck2001rubidium} , there has been so far no absolute data available for the transition 5S\,$\rightarrow$\,6P\,$\rightarrow$\,nS, nD, which is however necessary for the implementation of quantum information protocols using this excitation path. Knowing the 6P transition frequencies and the fine and hyperfine structure sub-levels with high accuracy, the transition frequencies from the 6P intermediate state to Rydberg states can then be calculated using the quantum defect theory \cite{mack2011measurement, PhysRevA.67.052502, PhysRevA.74.054502, PhysRevA.74.062712}.
Measurements of the hyperfine splitting of the $^{85}$Rb and $^{87}$Rb 6P levels have been performed before \cite{Arimondo77, Bize99}, but the most accurate value for the absolute frequencies in literature have uncertainties of \SI{850}{MHz} (\SI{0.0005}{\nano \metre}) \cite{NIST2018}.
Here, we report on the measurements of the absolute frequencies for this transitions and on measurement schemes to determine the relative frequencies. We have measured the absolute frequencies of the 5S\,$\rightarrow$\,6P$_{3/2}$ resonance with an uncertainty of \SI{\leq 20}{kHz} and better than \SI{450}{kHz} for the 5S\,$\rightarrow$\,6P$_{1/2}$ transition, improving the literature values by five and four orders of magnitude, respectively. 
In addition to verifying the literature values of the isotope shifts and the magnetic dipole and electric quadrupol constants, we have also determined the hyperfine splittings, depicted in Fig.\,\ref{abstaende}, from the measured transition frequencies, which brings their accuracies to the same respective orders of magnitude, improving the literature values by three and two orders of magnitudes.

\section*{Experimental Setup}
\begin{figure}[tbp]
	\centerline{
		\resizebox{9cm}{!}{
			\includegraphics[width=\linewidth]{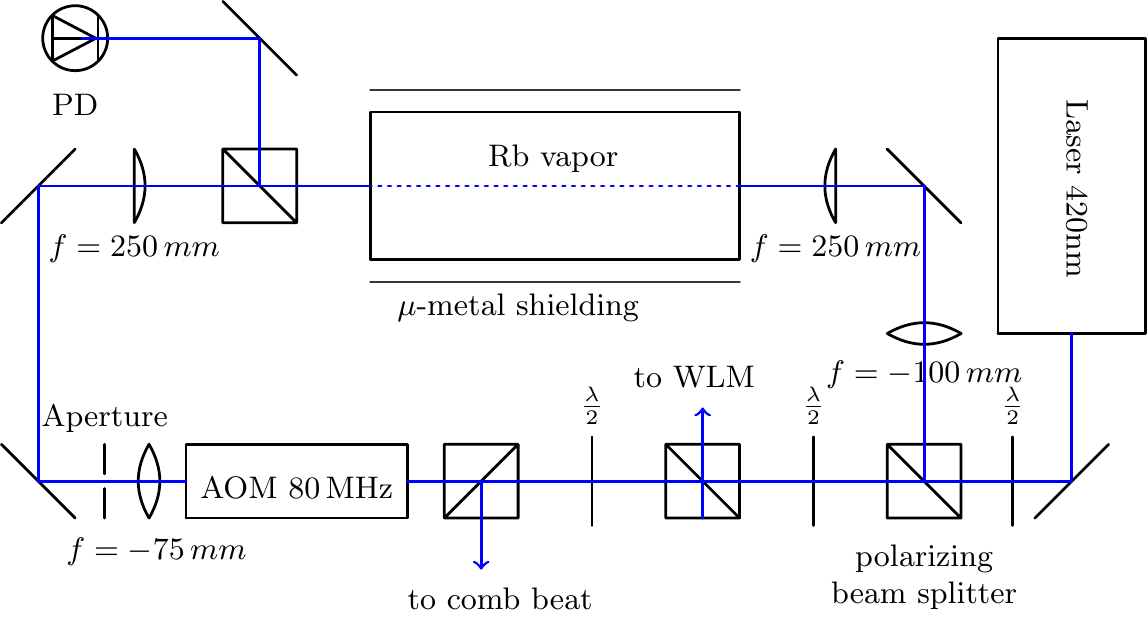}
	}}
	\caption{(Color online) Optical setup for the measurement of the 5S $\rightarrow$ 6P transitions in a rubidium vapor cell. The probe and pump beams are counter propagating in the cell (heated to \SI{\approx 321}{\K}). The probe laser intensity is monitored with a photodiode (PD) and the pump beam is chopped with an 80-MHz acousto-optical modulator (AOM). Both beams originate from an external cavity diode laser, which is either scanned by a wavelength meter (WLM) (5S $\rightarrow$ 6P$_{1/2}$ transition) or locked to the frequency comb (5s $\rightarrow$ 6P$_{3/2}$ transition).}
	\label{SSP}
\end{figure}

\begin{figure}[tbp]
	\centering
		\includegraphics[width=\linewidth]{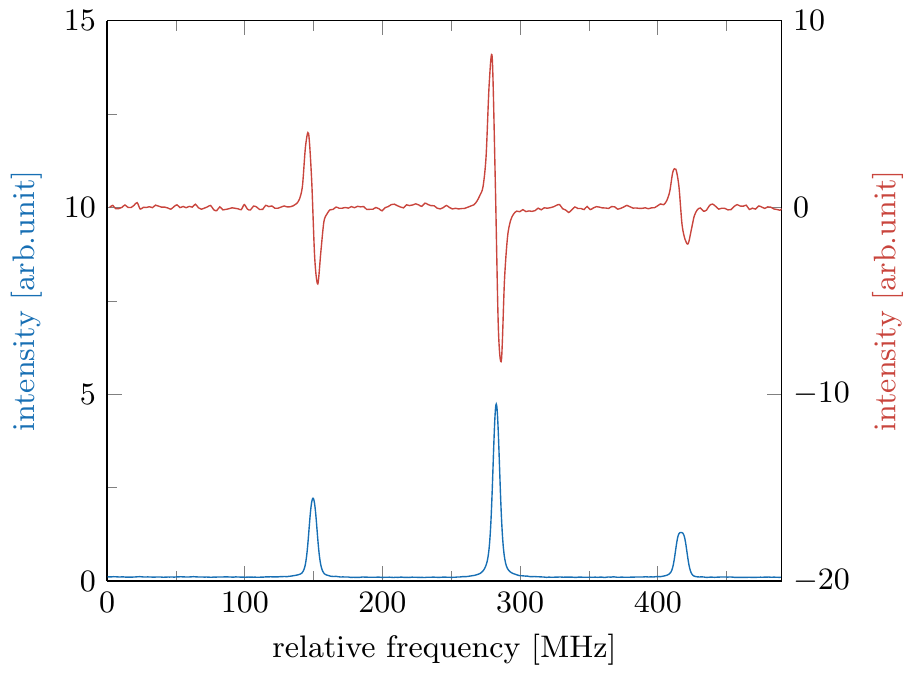}
	\caption{Doppler free spectrum (blue, color online) and the resulting error signals (red) for the  $^{87}$Rb 5S$_{1/2}$ (F=1) $\rightarrow$ 6P$_{1/2}$ transitions.}
	\label{Fehler}
\end{figure}

Our experimental setup is a saturation spectroscopy as shown in Fig.\,\ref{SSP}. 
The laser source is an ECDL with a linewidth of \SI{< 400}{kHz} as measured by the beating signal between the ECDL laser and the frequency comb. The frequency comb is phase-locked to a solid state \SI{1550}{\nano \metre} laser (I15, NKT) with a linewidth of $<\SI{100}{Hz}$. The linewidth of the comb is $<$\SI{2}{\kilo Hz}, which was measured by beating it to a second \SI{1550}{\nm} laser.
The pump and probe beams counter-propagate through the heated rubidium vapor cell (\SI{321}{\K}). 
Polarizing beam splitters ensure crossed linear polarizations to avoid interference effects within the cell. 
A double-layer magnetic $\mu$-metal shielding leads to a reduction of magnetic fields to less than \SI{0.03}{G}, which has been verified with a Gaussmeter (GM07 Gaussmeter, Hirst Magnetic Instruments LTD).
The power ratio between probe and pump beams was adjusted in order to optimize the signal-to-noise ratio of the Lamb dips.
The optimal ratio was found to be near (10:1), using a probe power of \SI{0.488}{mW} 
(\SI{14.9}{mW/cm^2}) and a pump beam power of \SI{56}{\uW} (\SI{1.2}{mW/cm^2}).  
The intensity and the frequency of the pump beam were modulated with a frequency of \SI{50}{kHz} using an \SI{80}{MHz} acousto-optical modulator (AOM). 
The probe beam signal from the photodiode was subsequently demodulated at the same frequency using a lock-in Amplifier (HF2LI, Zurich Instruments), which results in a fully Doppler free spectroscopy signal \cite{ye1996hyperfine}, as depicted exemplarily in Fig.\,\ref{Fehler} (blue line) for the $^{87}$Rb 5S$_{1/2} (F=1) \rightarrow$ 6P$_{1/2} (F'=1)$ transitions. Since the AOM shifts the pump beam by \SI{+80}{MHz} the measured Lamb dips and cross-overs are red-shifted by \SI{-40}{MHz}. 
This offset has been corrected in the final data analysis.\\
The relative frequencies between the 5S\,$\rightarrow$\,6P$_{1/2}$ transitions were measured 
by scanning the laser with a wavemeter and simultaneously recording the wavelength of the laser.

%
Additionally, we can impose \SI{3}{MHz} sidebands by frequency modulation (FM) of the laser diode current.
After appropriate demodulation, this leads to a Pound-Drever-Hall-like (PDH) error signal, which can be used to stabilize the laser onto the transition resonance frequency \cite{Drever1983}. 
Fig.\,\ref{Fehler} shows an exemplary spectrum acquired by the Lock-in amplifier (blue) and the FM error signals (red) for the  $^{87}$Rb 5S$_{1/2} (F=1) \rightarrow$ 6P$_{1/2}$ transitions.\\
We can control the frequency of the laser in three ways: 
The first is to stabilize the laser frequency with the FM error signal to one of the transitions using a digital laser locking module (Di\-gi\-Lock, TOPTICA).  
Alternatively, the laser frequency can be stabilized and scanned by the wavelength meter, which is calibrated with a laser that is frequency stabilized to the 5S$_{1/2}$ (F=2) $\rightarrow$ 5P$_{3/2}$ (F'=3) transition of $^{87}$Rb (\SI{780.246291}{\nano \metre}) via an FM spectroscopy. 
Those methods were used to measure the absolute and the relative frequencies for both transitions.
For the third method the beam of the ECDL and the modes of a narrow linewidth frequency comb (TOPTICA) were superimposed and frequency-filtered by a grating in a beat detection unit (DFC BC and DFC MD, TOPTICA) and monitored on a photodetector with a bandwidth of \SI{50}{MHz}. The resulting beating signal was acquired with a digital oscilloscope (Picoscope 5442A, Pico Technology). Using the beating signal the \SI{420}{\nano \metre} laser was phase-locked to the frequency comb, using a phase frequency detector (PFD, Toptica) with a filter, that is tunable between 2 and \SI{38}{MHz}. 
	\begin{figure}[tbp]
	\centering
		\includegraphics[width=\linewidth]{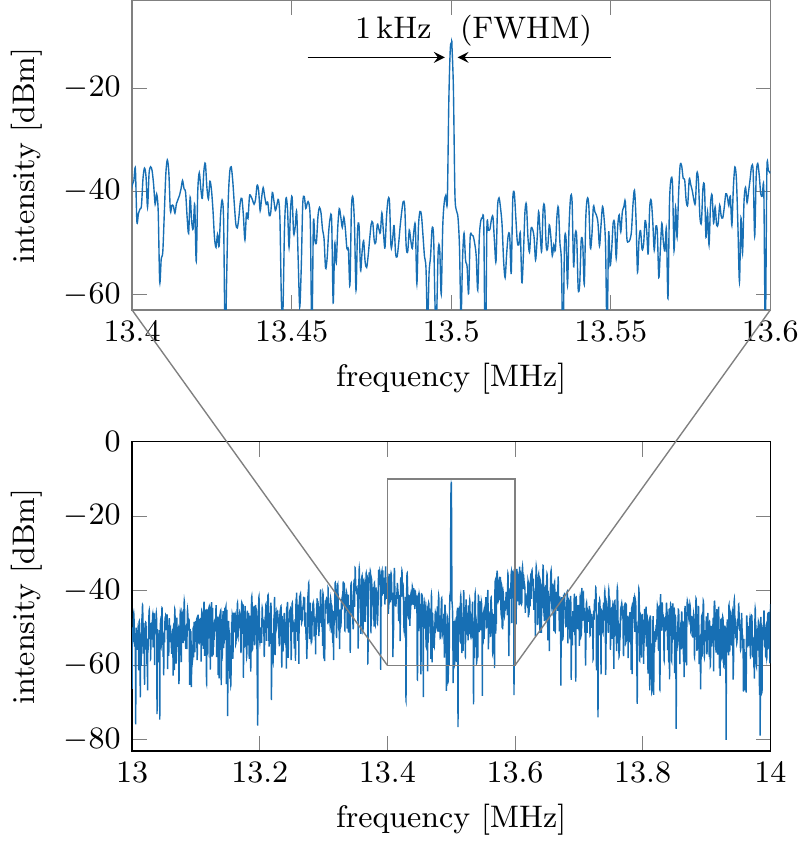}
	\caption{Typical phase-locked beating signal between a CW laser and a narrow line frequency comb. The upper graph is a zoom into the region of the beating signal, showing a width of \SI{\approx 1}{kHz}.}
	\label{Beat}
\end{figure}
A typical phase-locked beating signal with a width of \SI{\approx 1}{kHz} 
can be seen in Fig.\,\ref{Beat}.
%
%
The frequency comb is offset-free and its repetition rate $f_{\text{rep}}$=\SI{80}{MHz} is locked to a GPS-based \SI{10}{MHz} frequency reference \cite{Puppe:16} with an accuracy of $10^{-10}$. The comb is phase-locked to the \SI{1500}{\nano\m} laser using a phase frequency detector, which results in comb modes with a linewidth of $<\SI{5}{\kilo\Hz}$, measured by locking the comb to another frequency comb (FC 1500, Menlo). 
 This method to measure the frequencies was only applicable to the $5S \rightarrow$ 6P$_{3/2}$ transition, since the comb doesn't support the corresponding wavelength to the $5S \rightarrow$ 6P$_{1/2}$ transition. 
\begin{figure}[tbp]
	\centering
		\includegraphics[width=\linewidth]{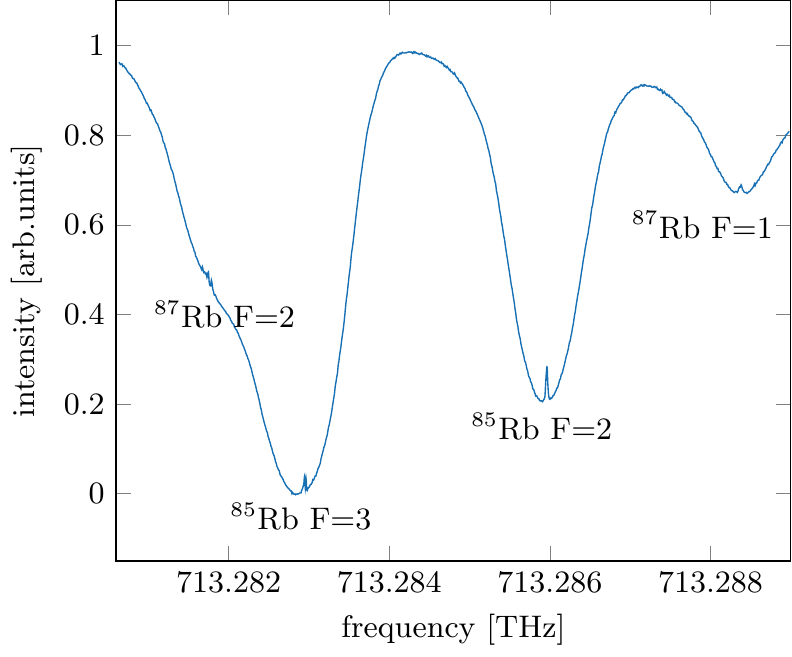}
	\caption{Saturation spectrum for the 5S $\rightarrow$ 6P$_{3/2}$ transitions for $^{85}$Rb (inner Dips) and $^{87}$Rb (outer dips).}
	\label{Spek}
\end{figure}
Fig.\,\ref{Spek} shows a typical saturation spectrum for the 5S $\rightarrow$ 6P$_{3/2}$ transition.

\section*{Measurement of the 5S $\rightarrow$ 6P$_{3/2}$ transitions}
For the measurement of the absolute transition frequencies the ECDL laser was beated with the frequency comb. The laser was locked to the FM error signal of each transition. Simultaneously, the wavemeter recorded the frequency and a digital oscilloscope saved the beat frequency. Knowing these values and the repetition rate of the comb, we were not only able to calculate the absolute frequencies, but also to determine the related comb mode. In order to reduce the statistical error, each measurement was repeated 60 times and subsequently averaged. To characterize the locking accuracy, we locked the \SI{420}{\nano \metre} laser to an arbitrary frequency for \SI{1}{h} and recorded the beating signal between laser and frequency comb every \SI{15}{s}. We found that the lock frequency of the laser deviates less than \SI{11.6}{kHz/h} from its mean value.\\
\begin{figure*}[t]
	\centering
	\includegraphics[width=\linewidth]{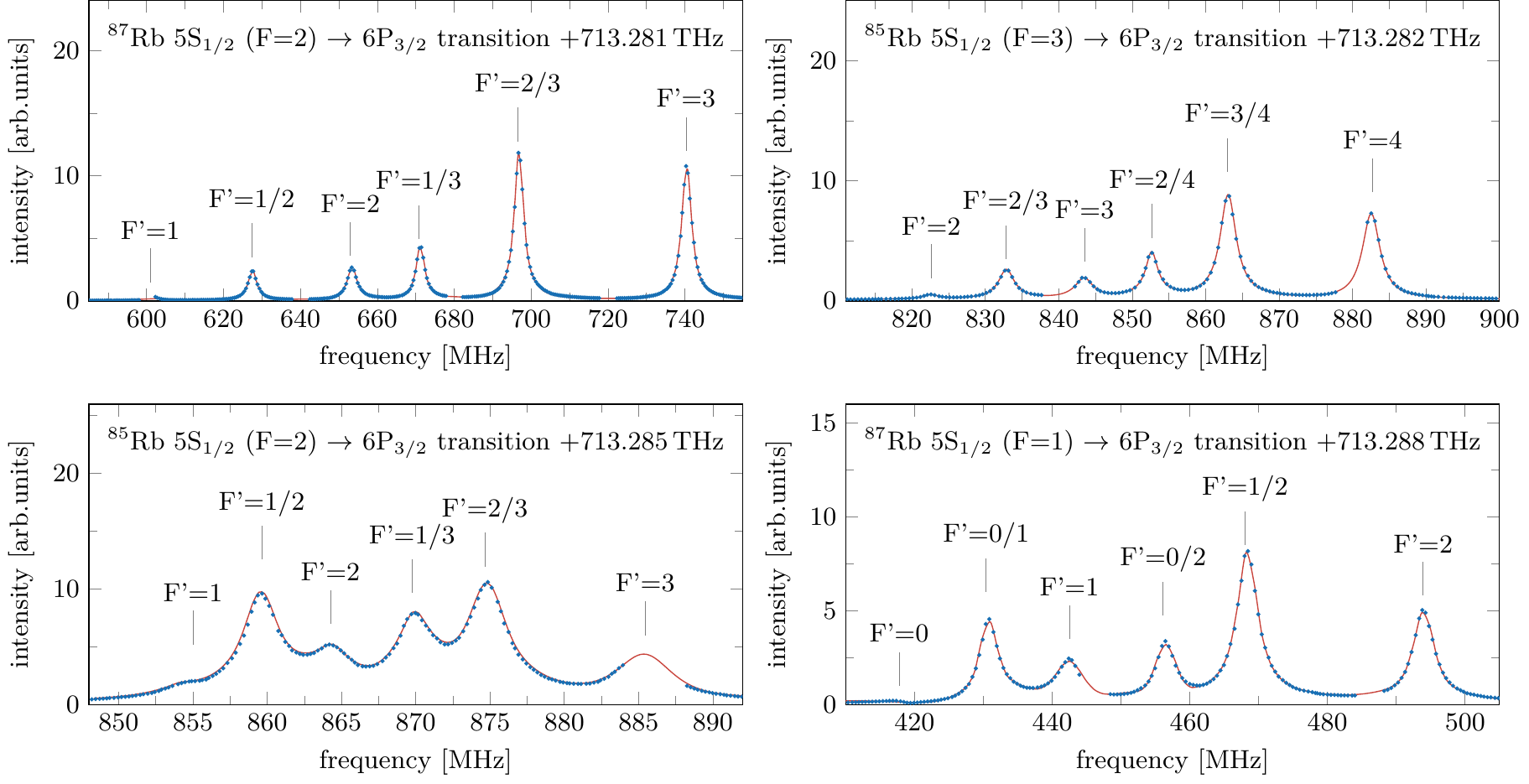}
	\caption{(Color online) Recorded spectra (blue dots) and fitted superpositions of Pseudo-Voigt functions (red solid line) for the 5S$_{1/2} \rightarrow$ 6P$_{3/2}$ transition. The gaps in the spectra are caused by the PFD locking scheme, which can only be tuned between 2 and 38\,MHz for each comb mode, resulting in 4\,MHz gaps whenever the comb mode has to be changed.}
	\label{spectren420}
\end{figure*}
The relative frequencies were determined by locking the laser to the frequency comb at a given locking point using the PFD. Starting at the previously measured frequencies and scanning this locking point between 2 and \SI{38}{MHz} for the corresponding comb mode, the relative frequencies of the 5S$_{1/2} \rightarrow$ 6P$_{3/2}$ transitions can be determined with an accuracy of \SI{<1}{kHz} resulting from the linewidth of the ECDL laser locked to the frequency comb. Figure \ref{spectren420} shows the resulting spectra from these scans, which have been calibrated using the absolute frequencies from the wavemeter measurement. The gaps in the measured spectra are caused by the locking scheme with the PFD, which can only be tuned between 2 and \SI{38}{MHz} for each comb mode, since there has to be a frequency difference to the next comb mode. This results in \SI{4}{MHz} gaps whenever the comb mode has to be changed. All spectra are fitted using a superposition of Pseudo-Voigt profiles, yielding a linewidth of \SI{\approx 2.7}{MHz} (FWHM). The errors arising from the fit routine was calculated separately for each spectrum and were found to be smaller than \SI{20}{kHz}.\\
The uncertainty in the data for the absolute frequency comprises several known error sources of physical and technical nature. The laser linewidth of the phase-locked laser (\SI{1}{kHz}), technical noise and physical deviations cause the fitting error to contribute to the uncertainty of the measured frequency. The overall uncertainty resulting from these effects was determined independently for each measured transition and was found to be \SI{\leq 20}{kHz}. The absolute transition frequencies are summarized in Tab. \ref{6p32}.\\
We have also performed measurements of the absolute transition spectra using the wavemeter to lock the laser. The experimental sequence was as follows: First, the wavelength meter was calibrated as described above. Second, the laser frequency was swept linearly at a rate of \SI{50}{MHz/\second} over a range of \SI{250}{MHz} while the signal of the lock-in amplifier was recorded. In order to reduce the statistical error, each trace has been measured 100 times and then averaged. The deviations between this measurement and the one including the frequency comb was found to be smaller than \SI{500}{kHz}. 
\begin{table}
	\centering
	\caption{Measured absolute transition frequencies between the 5S$_{1/2}$ and 6P$_{3/2}$ states of $^{85}$Rb and $^{87}$Rb.}
	\begin{tabular}{|l|l|c|}
		\hline
		& Transition & Frequency [THz]\\
		\hline
		$^{87}$Rb & F=2 $\rightarrow$ F'=1 & 713.281601641(16)\\
		& F=2 $\rightarrow$ F'=1/2 & 713.281627545(16)\\
		& F=2 $\rightarrow$ F'=2 & 713.281653455(16)\\
		& F=2 $\rightarrow$ F'=1/3 & 713.281671206(16)\\
		& F=2 $\rightarrow$ F'=2/3 & 713.281696845(16)\\
		& F=2 $\rightarrow$ F'=3 & 713.281740464(16)\\	
		\hline
		$^{85}$Rb  & F=3 $\rightarrow$ F'=2 & 713.282822529(16)\\
		& F=3 $\rightarrow$ F'=2/3 & 713.282832859(16)\\
		& F=3 $\rightarrow$ F'=3 & 713.282843436(16)\\
		& F=3 $\rightarrow$ F'=2/4 & 713.282852661(16)\\
		& F=3 $\rightarrow$ F'=3/4 & 713.282863062(16)\\
		& F=3 $\rightarrow$ F'=4 & 713.282882547(16)\\	
		\hline
		$^{85}$Rb& F=2 $\rightarrow$ F'=1 & 713.285854135(18)\\
		& F=2 $\rightarrow$ F'=1/2 & 713.285859657(18)\\
		& F=2 $\rightarrow$ F'=2 & 713.285864366(18)\\
		& F=2 $\rightarrow$ F'=1/3 & 713.285869902(18)\\
		& F=2 $\rightarrow$ F'=2/3 & 713.285874803(18)\\
		& F=2 $\rightarrow$ F'=3 & 713.285885399(18)\\
		\hline
		$^{87}$Rb& F=1 $\rightarrow$ F'=0 & 713.288419955(20)\\         
		& F=1 $\rightarrow$ F'=0/1 & 713.288430802(20)\\
		& F=1 $\rightarrow$ F'=1 & 713.288442281(20)\\
		& F=1 $\rightarrow$ F'=0/2 & 713.288456485(20)\\
		& F=1 $\rightarrow$ F'=1/2 & 713.288468381(20)\\
		& F=1 $\rightarrow$ F'=2 & 713.288493970(20)\\
		\hline
	\end{tabular}
	\label{6p32}
\end{table}

\section*{Measurement of the 5S $\rightarrow$ 6P$_{1/2}$ transitions}
\begin{figure*}
	\centering
		\includegraphics[width=\linewidth]{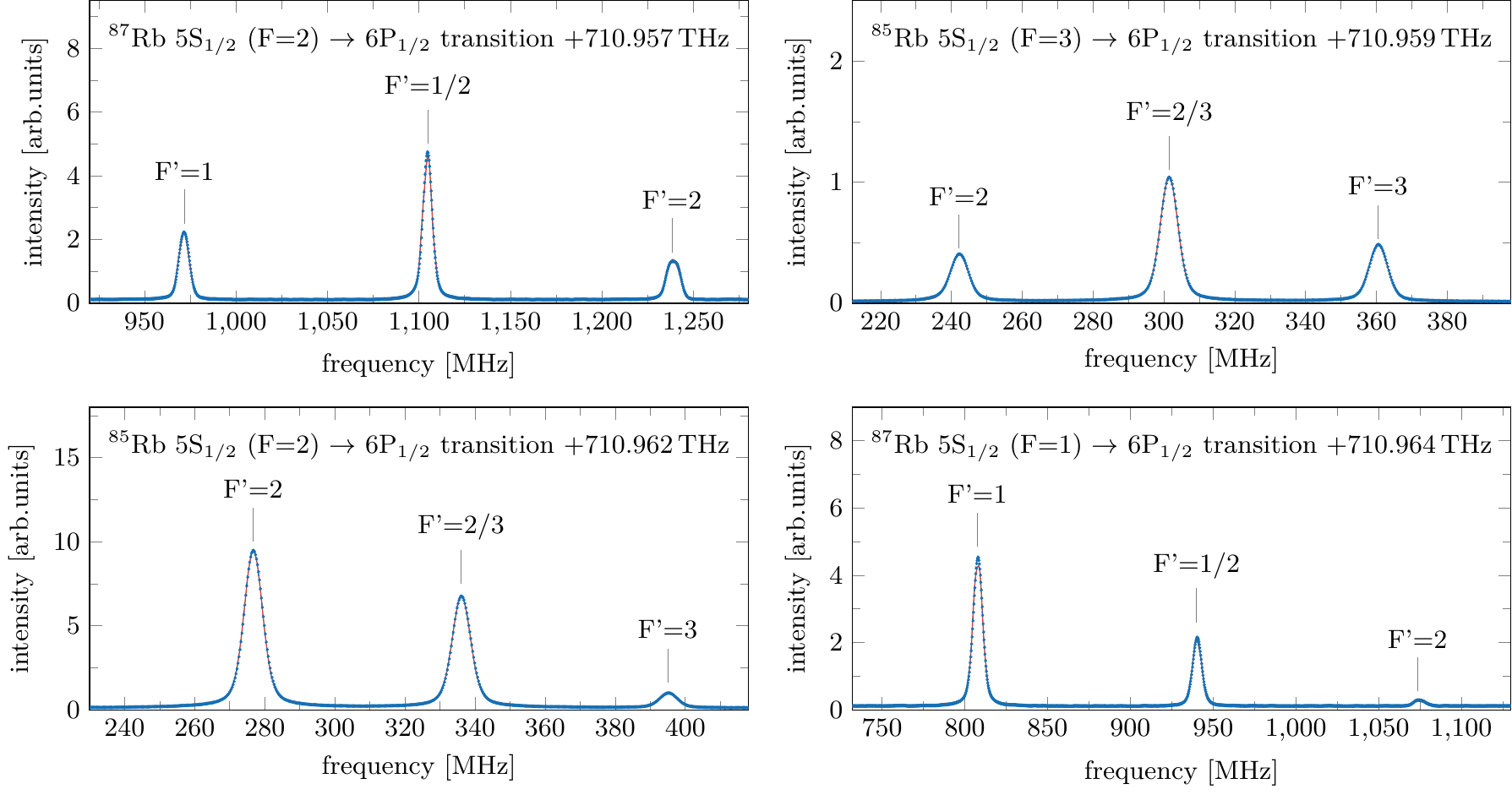}
	\caption{Recorded spectra (blue dots, color online) and fitted superpositions of Pseudo-Voigt functions (red solid line) for the 5S$_{1/2} \rightarrow$ 6P$_{1/2}$ transitions.}
	\label{spectren421nm}
\end{figure*}

%
Since the output of the frequency comb near \SI{420}{\nano \metre} is limited to a range between \SI{418.8}{\nano \metre} and \SI{420.3}{\nano \metre} it was not possible to obtain beating signals for the  5S$_{1/2}\,\rightarrow$\,6P$_{1/2}$ transition at \SI{421}{\nano \metre}. Therefore, the frequency measurements for this manifold are based on the DigiLock module and wavemeter scans. As determined for the 5S$_{1/2}\,\rightarrow$\,6P$_{3/2}$ transition, these methods show deviations of about \SI{500}{kHz} from each other, so we deem this to be the maximum error.
Similarly, as described above, the frequencies were determined by locking the laser to the error signal using the DigiLock. Then the laser was locked to a calibrated wavemeter and swept at a rate of \SI{50}{MHz/s} over a range of \SI{500}{MHz} to measure the spectra.
Again, each trace was measured 100 times and subsequently averaged, to reduce statistical errors. The resulting spectra are depicted in Fig.\,\ref{spectren421nm}. 
We typically measured linewidths (FWHM) of \SI{\approx 2.8}{MHz}, roughly twice as large as expected from the natural linewidth of \SI{1.135}{MHz} \cite{safronova2011critically}.
The positions of the peaks were evaluated by fitting superpositions of Pseudo-Voigt profiles to the spectra. The error due to the fitting routine is on the order of a few kHz and can be neglected compared to the deviations due to the linewidth of the laser. To estimate the error caused by the wavemter we have characterized its locking accuracy. Therefore we have locked the \SI{420}{\nano \metre} laser to an arbitrary frequency within the output range of the frequency comb for \SI{1}{h} and recorded the beat signal between laser and a frequency comb mode every \SI{15}{\second}. We found the frequency of the calibrated wavemeter to be normally distributed with a standard deviation of \SI{160}{kHz}. 
%
%
The total error of the frequencies was calculated to be $<$450\,kHz, combining the laser linewidth of \SI{< 400}{kHz} and the relative error of the callibrated wavemeter of $160$\,kHz. The results for the transition frequencies are summarized in Tab. \ref{6p12}.

\begin{table}
	\centering
	\caption{Measured absolute transition frequencies between the 5S$_{1/2}$ and 6P$_{1/2}$ states of $^{85}$Rb and $^{87}$Rb. }
	\begin{tabular}{|l|l|c|}
		\hline
		& Transition & Frequency [THz]\\
		\hline
		$^{87}$Rb& F=2 $\rightarrow$ F'=1 & 710.95797053(45)\\
		& F=2 $\rightarrow$ F'=1/2& 710.95810465(45)\\
		& F=2 $\rightarrow$ F'=2 & 710.95823892(45)\\
		\hline
		$^{85}$Rb& F=3 $\rightarrow$ F'=2 & 710.95924220(45)\\
		& F=3 $\rightarrow$ F'=2/3 & 710.95930147(45)\\
		& F=3 $\rightarrow$ F'=3 & 710.95936053(45)\\
		\hline
		$^{85}$Rb  & F=2 $\rightarrow$ F'=2 & 710.96227673(45)\\
		& F=2 $\rightarrow$ F'=2/3 & 710.96233612(45)\\
		& F=2 $\rightarrow$ F'=3 & 710.96239520(45)\\
		\hline
		$^{87}$Rb & F=1 $\rightarrow$ F'=1 & 710.96480730(45)\\         
		& F=1 $\rightarrow$ F'=1/2 & 710.96494083(45)\\
		& F=1 $\rightarrow$ F'=2 & 710.96507276(45)\\
		\hline
	\end{tabular}
	\label{6p12}
\end{table}


\section*{Discussion}
Since pressure broadening is negligible (\SI{< 1}{kHz}) and the Zeeman shift caused by the magnetic field is below \SI{10}{kHz}, the broadening of the natural linewidth discrepancy is assumed to be mostly the consequence of the transversal Doppler effect (1\,MHz/0.12$^\circ$) and power broadening. \\
With the measured data we calculated the hyperfine splitting for both transitions. The errors are calculated via propagation of uncertainty based on the transition frequency errors. In Tab. \ref{HF} a comparison between the literature values and the newly determined values is shown.\\
\begin{table}[tbp]
	\centering
	\caption{Hyperfine splitting. The literature values were given without errorbars.}
		\begin{tabular}{|l||r|r|}
			\hline
	Transition $^{85}$Rb & Frequency [MHz] \cite{grundevik1977high} & Frequency [MHz]\\
		\hline
6P$_{1/2}$ F=2 - F=3 & 117 & 118.40(46)\\
		\hline
6P$_{1/2}$  & 68.25 & 69.07(46)\\
		\hline
6P$_{3/2}$ F=1 - F=2 & 9 &  10.048(25)\\
		\hline
6P$_{3/2}$ F=2 - F=3 & 21 &  20.967(25)\\
\hline
6P$_{3/2}$ F=3 - F=4 & 39 &  39.127(25)\\
		\hline
6P$_{3/2}$ & 32.5 &  32.703(25)\\
		\hline
		\hline
	Transition $^{87}$Rb & Frequency [MHz] \cite{grundevik1977high} & Frequency [MHz]\\
		\hline
6P$_{1/2}$ F=1 - F=2 & 265 & 265.15(46)\\
		 \hline
6P$_{1/2}$ & 165.625 & 165.72(46)\\
		 \hline
6P$_{3/2}$ F=0 - F=1 & 24 &  23.739(26)\\
		 \hline
6P$_{3/2}$ F=1 - F=2 & 52 &  51.654(26)\\
		\hline
6P$_{3/2}$ F=2 - F=3 & 87 &  87.009(26)\\
		\hline
6P$_{3/2}$  & 63.331 & 63.340(26) \\
		\hline
	\end{tabular}
	\label{HF}
\end{table}
The total isotope shifts were found to be \SI{41.935(60)}{MHz} for the 6P$_{3/2}$ isotopes and \SI{41.237(62)}{MHz} for the 6P$_{1/2}$ isotopes, which agree well with the literature values \cite{grundevik1977high, aldridge2011experimental}.\\
With the determined hyperfine splitting values, the magnetic dipole constant A and the electric quadrupole constant B can be calculated for the 5S $\rightarrow$ 6P transitions. The values for each $^{85}$Rb und $^{87}$Rb are listed in Tab. \ref{AB} and are also in good agreement with the literature data.
\begin{table}[tbp]
	\centering
	\caption{Calculated values for the magnetic dipole constant A and the electric quadrupole constant B.}
	\begin{tabular}{|l||c|r|r|}
		\hline
		Transition & Constant & literature  [MHz] & this work [MHz]\\
		\hline
	$^{85}$Rb 5S $\rightarrow$ 6P$_{3/2}$ & A & 8.179(12) \cite{Arimondo77} & 8.174(119)\\
		\hline
	$^{85}$Rb 5S $\rightarrow$ 6P$_{3/2}$ & B & 8.190(49) \cite{Arimondo77} & 8.113(119)\\
	\hline
	$^{87}$Rb 5S $\rightarrow$ 6P$_{3/2}$ & A & 27.700(17) \cite{safronova2011critically} & 27.716(108)\\
	\hline	
	$^{87}$Rb 5S $\rightarrow$ 6P$_{3/2}$ & B & 3.953(24) \cite{safronova2011critically} & 3.983(108)\\
	\hline
	$^{85}$Rb 5S $\rightarrow$ 6P$_{1/2}$ & A & 39.11(3) \cite{Arimondo77} & 39.47(50)\\
	\hline
	$^{87}$Rb 5S $\rightarrow$ 6P$_{1/2}$ & A & 132.56(3) \cite{safronova2011critically} & 132.83(50)\\
	\hline
	\end{tabular}
	\label{AB}
\end{table}

\section*{Conclusion}
In summary, we have performed high precision saturated absorption spectroscopy of $^{85}$Rb and $^{87}$Rb using a diode laser. The laser was stabilized and scanned by a wavelength meter for the 6P$_{1/2}$ transition and locked to a narrow linewidth frequency comb for the 6P$_{3/2}$ transition. This allows for absolute frequency measurements with an uncertainty of \SI{< 20}{kHz} for the 6P$_{3/2}$ transition and \SI{< 450}{kHz} for the 6P$_{1/2}$ transition. The lower uncertainty in the measurement of the 5S $\rightarrow$ 6P$_{3/2}$ transition results from the laser locking to a narrow line frequency comb, while for the measurement of the 5S$ \rightarrow$ 6P$_{1/2}$ transition it is limited by the stability of the wavelength meter. From the measured data we derive hyperfine splitting values with unprecedented accuracy and verified the literature values for the isotope shifts, the magnetic dipole constant and the electric quadrupole constant.

\section*{Acknowledgements}
This work has been supported by the DFG (SSP 1929 GiRyd and CIT) and BMBF (FKZ: 13N14903). C.G. would like to thank the Evangelische Studienstiftung Villigst e.V.

\end{document}